\documentclass[10pt, conference]{IEEEtran}
\IEEEoverridecommandlockouts
\usepackage{cite}
\usepackage{amsmath,amssymb,amsfonts}
\usepackage{algorithmic}
\usepackage{graphicx}
\usepackage{textcomp}
\usepackage{xcolor}
\usepackage{colortbl}
\usepackage{array}
\usepackage{listings}
\usepackage{xcolor}
\usepackage{tabularx}
\usepackage{array}
\usepackage{fancyvrb}

\newcolumntype{P}[1]{>{\centering\arraybackslash}p{#1}}

\definecolor{delim}{RGB}{20,105,176}
\definecolor{numb}{RGB}{106, 109, 32}
\definecolor{string}{rgb}{0.64,0.08,0.08}

\lstdefinelanguage{json}{
    numbers=left,
    numberstyle=\small,
    frame=single,
    rulecolor=\color{black},
    showspaces=false,
    showtabs=false,
    breaklines=true,
    postbreak=\raisebox{0ex}[0ex][0ex]{\ensuremath{\color{gray}\hookrightarrow\space}},
    breakatwhitespace=true,
    basicstyle=\ttfamily\small,
    upquote=true,
    morestring=[b]",
    stringstyle=\color{string},
    literate=
     *{0}{{{\color{numb}0}}}{1}
      {1}{{{\color{numb}1}}}{1}
      {2}{{{\color{numb}2}}}{1}
      {3}{{{\color{numb}3}}}{1}
      {4}{{{\color{numb}4}}}{1}
      {5}{{{\color{numb}5}}}{1}
      {6}{{{\color{numb}6}}}{1}
      {7}{{{\color{numb}7}}}{1}
      {8}{{{\color{numb}8}}}{1}
      {9}{{{\color{numb}9}}}{1}
      {\{}{{{\color{delim}{\{}}}}{1}
      {\}}{{{\color{delim}{\}}}}}{1}
      {[}{{{\color{delim}{[}}}}{1}
      {]}{{{\color{delim}{]}}}}{1},
}

\def\BibTeX{{\rm B\kern-.05em{\sc i\kern-.025em b}\kern-.08em
    T\kern-.1667em\lower.7ex\hbox{E}\kern-.125emX}}
\begin{document}

\title{Secure and Trustful Cross-domain Communication with Decentralized Identifiers in 5G and Beyond}

\author{
    \IEEEauthorblockN{Hai Dinh-Tuan}
    \IEEEauthorblockA{
        \textit{Service-centric Networing} \\
        \textit{Technische Universität Berlin/T-Labs} \\
        Berlin, Germany \\
        hai.dinhtuan@tu-berlin.de
    }
    \and 
    \IEEEauthorblockN{Sandro Rodriguez Garzon}
    \IEEEauthorblockA{
        \textit{Service-centric Networking} \\
        \textit{Technische Universität Berlin/T-Labs} \\
        Berlin, Germany \\
        sandro.rodriguezgarzon@tu-berlin.de
    }
    \and 
    \IEEEauthorblockN{Jianeng Fu}
    \IEEEauthorblockA{
        \textit{Technische Universität Berlin} \\
        Berlin, Germany \\
        jianengf@gmail.com
    }
}

\newcommand{\ts}{\textsuperscript}

\maketitle

\begin{abstract}

In the evolving landscape of future mobile networks, there is a critical need for secure and trustful communication modalities to support dynamic interactions among core network components of different network domains. This paper proposes the application of W3C-endorsed Decentralized Identifiers (DIDs) to establish secure and trustful communication channels among network functions in 5G and subsequent generations. A new communication agent is introduced that integrates seamlessly with 5G-standardized network functions and utilizes a DID-based application layer transport protocol to ensure confidentiality, integrity, and authenticity for cross-domain interactions. A comparative analysis of the two different versions of the DID-based communication protocol for inter network function communication reveals compatibility advantages of the latest protocol iteration. Furthermore, a comprehensive evaluation of the communication overhead caused by both protocol iterations compared to traditional TCP/TLS shows the benefits of using DIDs to improve communication security, albeit with performance loses compared to TCP/TLS. These results uncover the potential of DID-based communication for future mobile networks but also point out areas for optimization.

\end{abstract}


\section{Introduction}

As a result of cooperative operations among mobile network operators and the many other stakeholders of a 5G ecosystem, public land mobile networks have undergone a considerable transformation towards multi-domain networks driven by network function virtualization \cite{yousaf2017nfv} and initiatives such as OpenRAN. This evolution enhances interoperability through open standards, while also emphasizing the need for robust access control to mitigate unauthorized access and cyber threats. The current use of TLS with X.509 certificates for mutual network function (NF) authentication in the 5G's Service-based Architecture (SBA) is overseen by centralized Certificate Authorities (CAs), posing security risks as evidenced by incidents like the DigiNotar breach \cite{prins2011diginotar}. The reliance on a single global CA for issuing X.509 certificates is increasingly impractical and insecure due to various technical, geopolitical, and other constraints \cite{GSMAssociation.2021}. An alternative proposed by the GSMA involves integrating individual CAs for each domain and establishing cross-certification agreements \cite{GSMAssociation.2021, jorquera2022design}. Despite not relying on a commonly trusted CA, cross certification adds complexity, becomes less economically feasible and secure with the growing number of stakeholders in future mobile networks. 

This paper discusses a novel approach to enhance cross-domain security in 5G core networks by the introduction of decentralized identity and key management for NFs within the control plane, as recently mentioned in \cite{DIDSSI6G} and \cite{garzon2024beyond}. It empowers NFs to establish secure communication channels, guaranteeing confidentiality, integrity, and authenticity without relying on a global CA or the need for cross-certification of CAs. By equipping NFs with Decentralized Identifiers (DIDs) \cite{WorldWideWebConsortium.822021}, NFs become able to share public keys for the establishment of secure channels securely and trustfully across network domains via a Verifiable Data Registry (VDR). The proposed approach differs to today's centralized identity and key management in 5G in the sense that it allows NFs to self-issue and self-manage identifiers while sharing the associated public key material in a decentralized fashion, e.g., via a commonly operated and governed distributed ledger. The latter acts hereby as a common source of truth for verification material that is crucial for secure and trustful networking in multi-stakeholder and zero-trust environments. The proposed approach is the first attempt to implement a 5G-tailored communication agent that makes use of the DIDComm application layer transport protocol \cite{DIDCommAries} for the communication among NFs. A detailed qualitative and quantitative comparison of the two available DIDComm versions with the TCP/TLS protocol stack is given. The analysis assesses the performance and security of the 2\ts{nd} generation of DIDComm compared to its predecessor and TCP/TLS. The aim of this work is to simplify and enhance the cross-domain security in 5G and beyond.

\section{Related work}

In today's mobile core networks, TCP/TLS, VPN, and IPSec are used to ensure confidentiality, integrity, and authenticity at different layers, despite their lower performance compared to state-of-the-art communication protocols. Additionally, their reliance on centralized certificate management across network domains is complex and prone to cyber threats. This section explores initiatives focused on enhancing communication performance and refining cross-domain identity and key management for improved security and efficiency, leveraging decentralization principles.

\subsection{Performance and Security}

QUIC \cite{iyengar2021rfc} utilizes UDP for quick, secure web connections, integrating TLS 1.3 to reduce latency and overhead associated with TLS 1.2. It enables secure connections with zero round trips and addresses the head-of-line blocking problem through multiplexing, outperforming traditional TCP/TLS. TwinPeaks \cite{cho2020twinpeaks} is a certificateless public key system, using cryptography that bars the Key Generation Center from accessing private keys. It links public keys to network parameters to prevent spoofing and simplifies management with a DNS-like structure, improving security, reducing energy, and cutting latency.



Dahlmanns et al. \cite{dahlmanns2022missed} report a low 6.5\% TLS adoption in (I)IoT, due to long-life industrial devices, outdated tech, and resource limits. Addressing this, Varo et al. \cite{varo2022dynamic} propose a TLS extension for energy-efficient devices that dynamically adjusts handshakes and encryption based on battery and security requirements, yielding up to 57.1\% energy savings and 9.4\% battery conservation. Others propose new cryptographic protocols like the Ephemeral Diffie-Hellman over CBOR Encoded Message Syntax (COSE) \cite{selander2021ephemeral}, which offers a secure method for key exchange over an insecure channel. It involves entities generating and encoding temporary key pairs per session, using the COSE format. These keys offer forward secrecy and secure the shared secret key's confidentiality and integrity. Jan et al. \cite{jan2022selwak} introduce SELWAK, a lightweight, anonymous authentication and key establishment protocol in vehicular networks, utilizing XOR operations and one-way hash functions. Theodore et al. \cite{theodore2021novel} propose ITESLA-CF for secure vehicle-to-road-side unit communication, focusing on lightweight mutual authentication and strong anonymity.

\subsection{Decentralized Identity and Key Management}

A DID uniquely references an entity and resolves to a \textit{DID document} \cite{WorldWideWebConsortium.822021}. Formatted as \texttt{did:<method>:<subject>}, the \textit{DID method} sets resolution rules, and the \textit{DID subject} uniquely identifies the entity. The DID document contains non-identity-revealing public keys linked to the entity referred to, with the entity holding the corresponding private keys to verify DID ownership. This enables a DID owner to pseudo-anonymously authenticate towards others, similar as with a self-issued, pseudo-anonymous X.509 certificate known from today's centralized public key infrastructures. A VDR guarantees hereby that there exists at any point in time only one valid DID document associated with a DID. Hence, a VDR securely stores DID documents and ensures that only the DID owner or an entity explicitly chosen by the DID owner is allowed to modify the DID owner's DID document in the VDR while all others with access to the VDR can read the latest version of the DID document. A VDR can either be realized as a centralized or a decentralized repository, with distributed ledgers being the most prominent incarnation of decentralized VDRs. In a distributed ledger-based VDR, the ledger becomes a common source of truth for DID documents because it can be commonly operated and equally governed by multiple stakeholders and technically links the DID document of an entity with the entity's DID in a secure and trustful manner. As a consequence, a DID and its associated document is fully self-issued and self-managed but pseudo-anonymous.

To enable non-anonymous identification with DIDs, the W3C introduced the \textit{Verifiable Credentials (VC)} model \cite{WorldWideWebConsortium.05.11.2021}. VCs are tamper-resistant containers of identity claims. They are cryptographically linked to a DID and they are digitally signed by a trusted 3\ts{rd} party. During the issuance process, a VC is handed over from the trusted 3\ts{rd} party to the DID owner. With a VC, a DID owner can proof to others that it owns tamper-proof identity claims that are attested by a trustful 3\ts{rd} party. The verification of the VC by others can then be conduced by accessing the DID document of the issuer in the VDR and using the contained public keys to verify the VC's digital signature of the issuer (trusted 3\ts{rd} party).

\textit{DIDComm} \cite{DIDCommAries} is a simplex and message-based application-layer transport protocol that utilizes DIDs and DID documents to establish and maintain secure and trustful end-to-end communication channels between DID owners. It ensures message authenticity and integrity as well as message encryption for confidentiality. This connectionless communication protocol is transport-agnostic and can therefore operate on top of various communication technologies such as TCP, WebSockets, HTTP/1, HTTP/2, Bluetooth or even NFC.

\section{Concept}
\label{sec:concept}

\begin{figure}
    \centering
    \includegraphics[width=0.8\columnwidth]{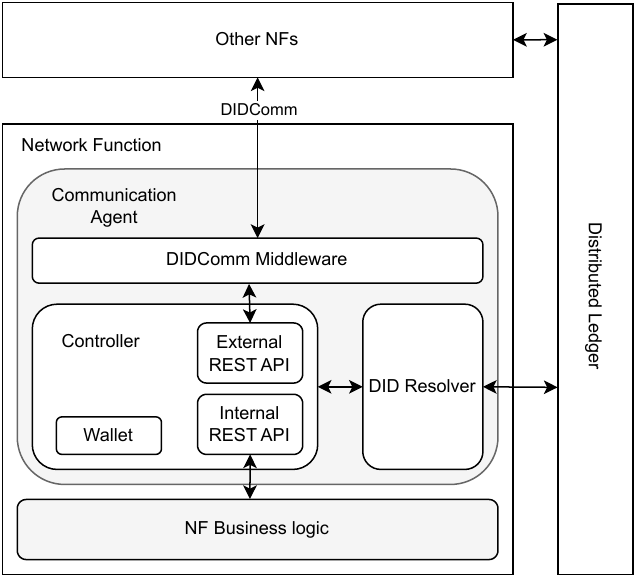}
    \caption{Integration of communication agent into each NF.}
    \label{fig:arch}
\end{figure}

\begin{figure}
	\centering
        \includegraphics[width=\columnwidth]{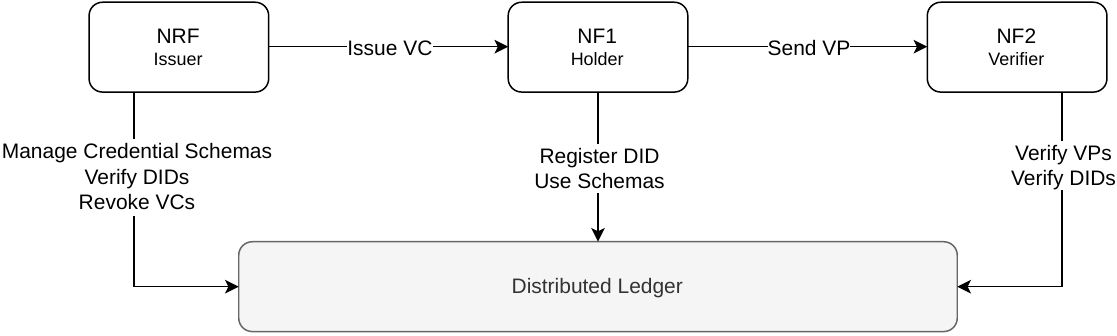}
	\caption{Authorization with the use of VCs.}
	\label{fig:authVC}
\end{figure}

The evolution from 1G/2G to 5G networks marks a significant shift towards a data-centric approach, introduces the \textit{Service-Based Architecture} where NFs are interconnected via standardized APIs over HTTP/2. This concept proposes using DIDs to establish secure communication channels among NFs with DIDComm and VCs for sharing identification information and permissions grants in a secure and trustful manner, replacing the traditional use of TLS with X.509 certificates for authentication and OAuth2.0 for authorization in the SBA. As depicted in Figure~\ref{fig:arch}, each NF integrates a DIDComm agent, separate from its core functionality, to manage DIDs, DIDComm, and VCs. The agent consists of DIDComm middleware, a DID resolver, and a controller with internal and external REST APIs for wallet management and message transmission. This implementation of DIDComm and VCs for NF authorization, as suggested by \cite{DIDSSI6G}, suggests replacing JSON Web Tokens (JWTs) with VCs issued by the Network Repository Function (NRF), which are then transmitted via DIDComm. The NRF, serving as a trusted issuer, validates these VCs as access tokens, thereby making the use of OAuth redundant. Nevertheless, as shown in Figure ~\ref{fig:authVC}, a VC schema must be first established in the ledger. The NRF issues VCs based on this schema and can revoke them using a revocation registry. NF1 sends its NRF-signed VC as a Verifiable Presentation (VP) to NF2 via DIDComm, which NF2 verifies using the issuer's and NF1's DID documents in the ledger.



\begin{figure}
	\centering
        \includegraphics[width=\columnwidth]{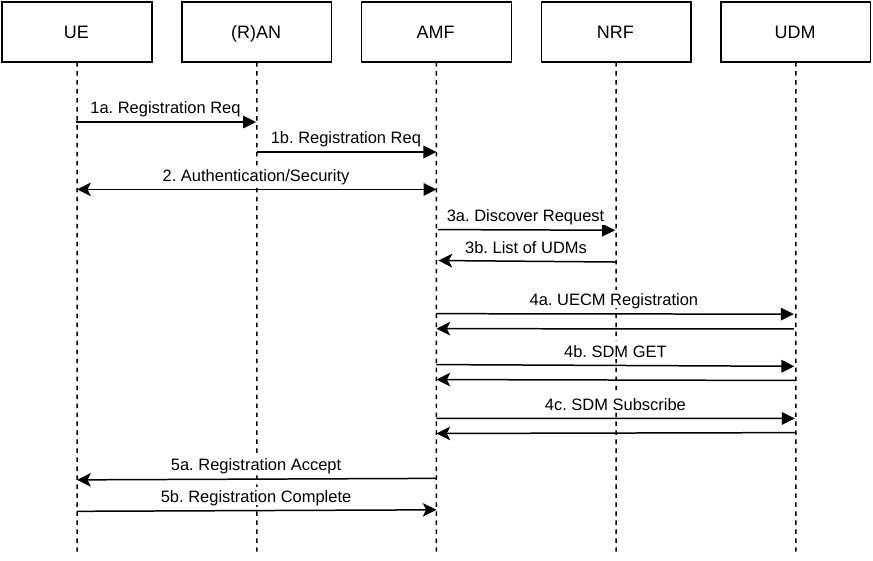}
	\caption{UE Registration as defined in 5G Release 17 \cite{ETSI}.}
	\label{fig:ueRegistration}
\end{figure}
\begin{figure}
	\centering
        \includegraphics[width=\columnwidth]{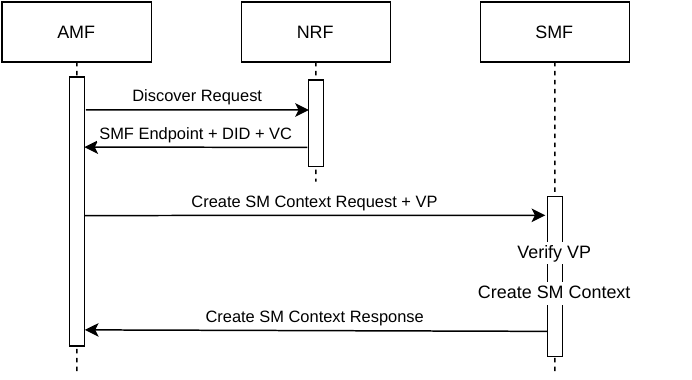}
	\caption{AMF requesting SM Context from AMF.}
	\label{fig:smContext}
\end{figure}

This paper delves into two key procedures defined in 5G Release 17 \cite{ETSI}, demonstrating the replacement of TCP/TLS with DIDComm for authentication, and OAuth2 with Present Proof protocol for authorization. The first scenario examines the User Equipment (UE) registration procedure, as depicted in Figure \ref{fig:ueRegistration}, with a focus on steps 3 and 4. Here, we propose the adoption of DIDComm over traditional TCP/TLS methods for authentication for NF-to-NF communications. In the second scenario, we explore a situation where the Access and Mobility Management Function (AMF) requests a Protocol Data Unit (PDU) session from the Session Management Function (SMF). This process commences with the AMF requesting a PDU session through a session management context from the SMF. Conventionally, the SMF requires AMF authorization, typically achieved via the OAuth2 protocol. Our proposed method, as illustrated in Figure~\ref{fig:smContext}, involves the Network Repository Function (NRF) providing the AMF with a JWT in the form of a VC for SMF authorization. The SMF then verifies this JWT by referencing the DID documents of both the NRF (issuer) and the AMF (holder).

\section{Evaluation}

For an experimental setup, diverse NFs of the 5G SBA, with each making use of DIDComm or TCP/TLS to interact with one another, were prototypically implemented to utilize either DIDComm or TCP/TLS for interactions, aiming to assess the feasibility of DIDComm as an alternative to TCP/TLS in 5G core networks. This setup, as described in Section \ref{sec:concept}, incorporates communication agents within each NF, integrating both DIDComm v1 and v2 due to their distinct features. The paper first conceptually examines the differences between these DIDComm versions, followed by a quantitative comparisons. Subsequently, we measure and compare the time required for secure message transmission among NFs using TCP/TLS, DIDComm v1, and v2. The final part of the analysis focuses on identifying which aspects of DIDComm message processing most significantly influence communication latency.




\subsection{Comparison of DIDComm v1 and v2}

\begin{table}
\centering
\def\arraystretch{1.2} 
\small 
\begin{tabular}{ | p{2.2cm} | p{2.5cm} | p{2.8cm} | } 
\rowcolor{gray!50}
\hline
\textbf{Property} & \textbf{DIDComm v1} & \textbf{DIDComm v2} \\
\hline
Project/author & Hyperledger Aries \cite{DIDCommAries} & Decentralized Identity Foundation \\
\hline
Interoperability & Aries-based & Broad interoperability \\
\hline
JSON encryption & Loosely based on JWE standard & JOSE standards \\
\hline
Message format & Custom format, not JWM compatible & Based on JWM \\
\hline
Connection setup & DID Exchange protocol & Protocol-independent \\
\hline
Messages headers & No message headers & At least 4 headers per message \\
\hline
DID method & Special handling for Peer DIDs & Uniform handling for all methods \\
\hline
\end{tabular}
\caption{Key differences between DIDComm v1 and v2}
\label{tab:didcomm}
\end{table}

DIDComm v1 and v2 differ in some key aspects impacting their suitability as TCP/TLS alternatives in 5G networks, summarized in Table \ref{tab:didcomm}.  Developed initially under the Hyperledger Aries project\footnote{https://www.hyperledger.org/projects/aries}, DIDComm v1 is limited to Aries agents. In contrast, DIDComm v2, now taken over by the Decentralized Identity Foundation (DIF), aims for broader interoperability and standardization through a Request for Comments (RFC). It uses standardized JSON Web Envelopes (JWEs) and encryption, eliminating the DID Exchange protocol for connection setup. DIDComm v2 doesn't require a prior connection setup, as peers can exchange DIDs and key material out-of-band or access updated DID documents via a distributed ledger. This simplifies processes but requires more message headers. While v1 uses a custom format not fully compatible with JSON Web Messages (JWM), v2 fully adopts JWM. DIDComm v2, less reliant on the Hyperledger Aries framework and more standardized, appears more promising for 5G networks. Both v1 and v2 are implemented in our prototype for evaluation, using different libraries: Hyperledger Aries project's ACA-Py agent\footnote{https://github.com/hyperledger/aries-cloudagent-python} for v1 and the Veramo project's agent\footnote{https://veramo.io/} for v2, due to each supporting only one DIDComm version.


\subsection{Message Size}

DIDComm v1 utilizes the DID Exchange protocol for establishing a communication association between peers, rendering the protocol stateful. This design minimizes the information required in each v1 message to identify its association with a specific communication link. In contrast, DIDComm v2 operates without a predefined communication association, necessitating each message to carry additional metadata. This metadata enables a receiving peer to process and associate the message with a sender independently, without maintaining the state of the communication link. However, this inclusion of extra metadata in DIDComm v2 messages results in a larger cumulative data exchange over time.

\begin{figure}
	\centering
    \includegraphics[width=0.8\columnwidth]{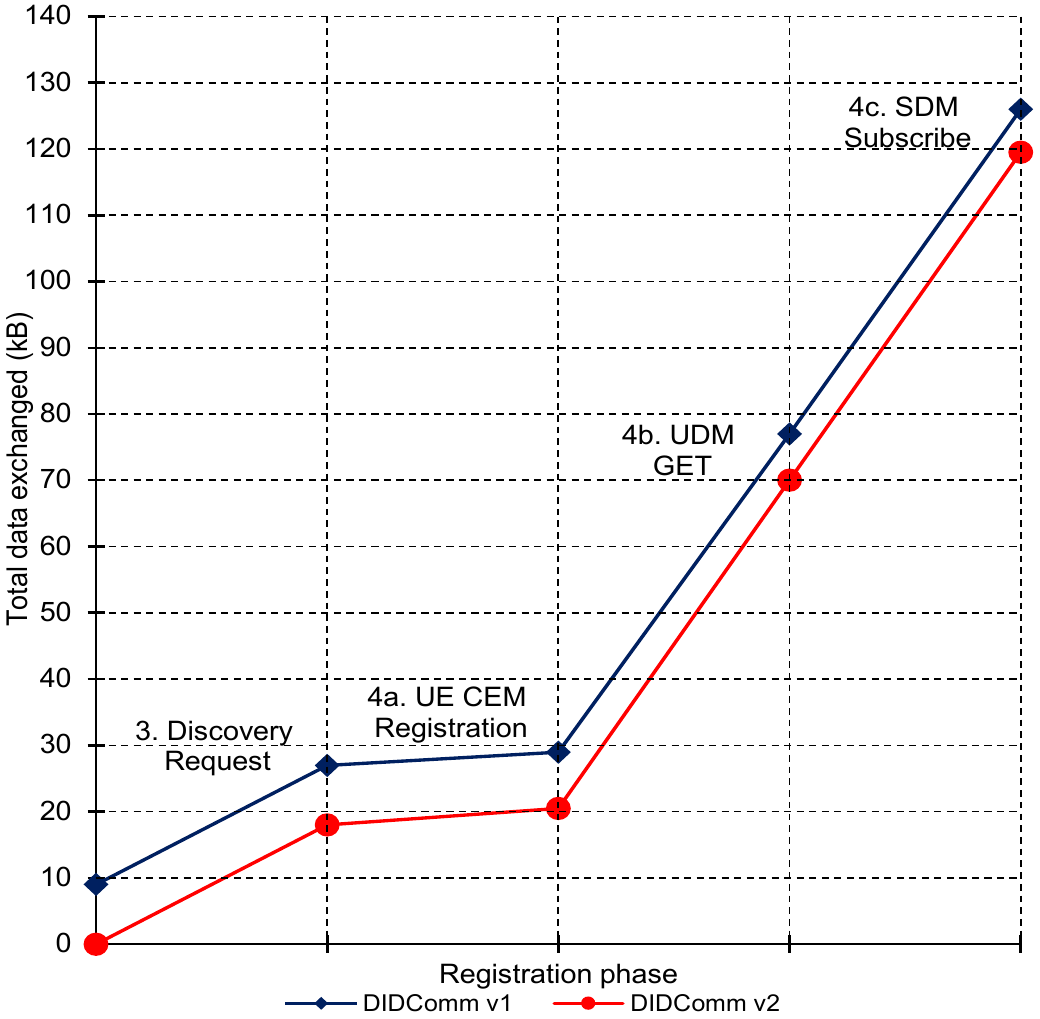}
	\caption{Cumulative bytes transmitted per UE registration step with DIDComm v1 and v2.}
	\label{fig:requests_0_4}
\end{figure}

Figure~\ref{fig:requests_0_4} shows the cumulative bytes transmitted per UE registration step, encompassing the SBA messages of step 3 and 4 of the overall UE registration process illustrated in Figure \ref{fig:ueRegistration}. Initially, DIDComm v1 transmits more data owing to the DID Exchange protocol previously discussed. Even at the conclusion of the UE registration process, the total data transmitted via DIDComm v2 remains lower than that transmitted via DIDComm v1. However, it is anticipated that, with ongoing usage of the same channel, the cumulative bytes transmitted through DIDComm v2 will eventually exceed those transmitted through DIDComm v1, primarily due to the smaller header size in DIDComm v1 messages.


\subsection{Latency}

In assessing latency, key phases such as encapsulation, decapsulation, and network transmission were evaluated for both DIDComm v1 (Python-based ACA-Py agent) and v2 (JavaScript-based Veramo agent). Both implementations used a local Docker deployment of Hyperledger Indy\footnote{https://www.hyperledger.org/projects/hyperledger-indy} as a VDR for storing and accessing DID documents.

\begin{figure}
        \includegraphics[width=1.015\columnwidth]{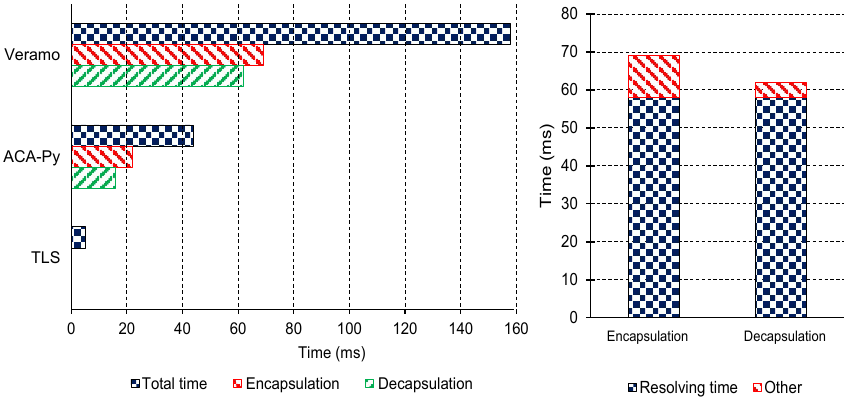}
	\caption{Duration for transmitting a message with DIDComm and TLS \& Analysis of encapsulationa and decapsulation processes with Verammo-based agent in DIDComm v2}
	\label{fig:compareTime}
\end{figure}

The evaluation measured the time from when the communication agent receives a request from the NF business logic to send an NF request via DIDComm, up to the point when the recipient's business logic can access the request content. Additionally, this processing time is compared to the latency experienced when using a TCP/TLS connection, as specified in 3GPP TS 33.501. The experiments with all versions, including and excluding the communication agents, were conducted on a machine with an i7-8550U CPU, 8GB of memory, and Ubuntu 16.04 LTS (kernel: 4.15.0-142-generic). Figure~\ref{fig:compareTime} presents the average outcomes for ten iterations for each agent variant. To realistically represent communication, the average message size involved in the UE registration procedure was calculated, yielding a payload size of 23,643 bytes.

DIDComm v2 communication showed an average latency of 158ms, with 69ms and 62ms for encapsulation and decapsulation, respectively, and about 27ms due to network latency and other processes. DIDComm v1, in contrast, had a lower latency of 44ms. Unlike TCP/TLS, which establishes a symmetric key only once, DIDComm's process necessitates digital signing, verification, and key derivation for every individual message. This connectionless nature of DIDComm explains its higher latency compared to TCP/TLS's 5ms average. However, in environments in which connection breaks are common, e.g., in case the SBA is extended to encompass the access network and even NFs executed within UEs itself, then a connectionless protocol as DIDComm might outperform a connection-oriented transport protocol such as TCP in conjunction with TLS.  

Despite similar processing steps for DIDComm message handling, ACA-Py significantly outperforms Veramo due to differences in resolving DID documents. The Veramo agent resolves the sender's and receiver's DID documents for each message without a cache, a process takes average 29ms per document after ten iterations with a locally-deployed Indy ledger. The network latency is therefore neglectable while the IndyDIDResolver's read operation (excluding the REST request processing and response generation) consumes an average of 14ms. Since a sender does not only resolve the DID document of the receiver before encapsulation a message but also its own one, the time doubles to an average of 58ms. The same applies for the decapsulation at the receiver.  

In contrast, the ACA-Py agent, using DIDComm v1, establishes a connection association with the DID Exchange protocol during setup, querying public DID documents only once and reusing them for subsequent communications. This approach, similar to a session concept, reduces message size and overall latency. The analysis of encapsulation and decapsulation process of the Veramo-based agent in Figure \ref{fig:compareTime} illustrates how DID document resolution in Veramo impacts latency, accounting for approximately 84\% during encapsulation and 94\% during decapsulation. From a conceptual point of view, caching a DID document with DIDComm v2 at each communication agent for a limited time span is also possible as long as the security policies permit it. However, in a real world setup, in which a copy of the distributed ledger is not expected to be available at each NF, the network latency of the DID resolution will have a non-neglectable impact on the overall latency.

\section*{Conclusion}

This paper aimed to address the administrative and security challenges of cross-domain authentication and authorization in the 5G core network by introducing a messaging agent based on the DIDComm v2 protocol, which can securely send and receive messages, facilitating credential exchanges for enhanced cross-domain trust. Our prototype, developed with Veramo and Hyperledger Indy, demonstrates that DIDComm can operate in 5G networks, reducing potential failure points and vulnerabilities. Comparing DIDComm v1 and v2, we found the latter to be more adaptable. The design of v2 explicitly aims for compatibility, utilizing open standards to ensure broader integrability across various systems. This makes DIDComm v2 not only more flexible but also more versatile for diverse applications. However, DIDComm’s latency, especially in our Veramo-based agent, is higher than TLS due to DID resolution, underscoring the need for performance improvements in DIDComm v2 without weakening security. Until then, its practicality remains uncertain. Future work should optimize resolution (e.g., caching) and test across more 5G procedures to validate this approach.

\bibliographystyle{IEEEtran}
\bibliography{bibliography}

\begin{thebibliography}{10}
\providecommand{\url}[1]{#1}
\csname url@samestyle\endcsname
\providecommand{\newblock}{\relax}
\providecommand{\bibinfo}[2]{#2}
\providecommand{\BIBentrySTDinterwordspacing}{\spaceskip=0pt\relax}
\providecommand{\BIBentryALTinterwordstretchfactor}{4}
\providecommand{\BIBentryALTinterwordspacing}{\spaceskip=\fontdimen2\font plus
\BIBentryALTinterwordstretchfactor\fontdimen3\font minus \fontdimen4\font\relax}
\providecommand{\BIBforeignlanguage}[2]{{%
\expandafter\ifx\csname l@#1\endcsname\relax
\typeout{** WARNING: IEEEtran.bst: No hyphenation pattern has been}%
\typeout{** loaded for the language `#1'. Using the pattern for}%
\typeout{** the default language instead.}%
\else
\language=\csname l@#1\endcsname
\fi
#2}}
\providecommand{\BIBdecl}{\relax}
\BIBdecl

\bibitem{yousaf2017nfv}
F.~Z. Yousaf, M.~Bredel, S.~Schaller, and F.~Schneider, ``{NFV and SDN—Key technology enablers for 5G networks},'' \emph{IEEE Journal on Selected Areas in Communications}, vol.~35, no.~11, pp. 2468--2478, 2017.

\bibitem{prins2011diginotar}
J.~R. Prins and B.~U. Cybercrime, ``{DigiNotar Certificate Authority breach “Operation Black Tulip”},'' \emph{Fox-IT, November}, vol.~18, 2011.

\bibitem{GSMAssociation.2021}
\BIBentryALTinterwordspacing
{GSM Association}, ``{Key Management for 4G and 5G Inter-PLMN Security},'' 2021, accessed: 2023-10-25. [Online]. Available: \url{https://www.gsma.com/security/resources/fs-34-key-management-for-4g-and-5g-inter-plmn-security/}
\BIBentrySTDinterwordspacing

\bibitem{jorquera2022design}
J.~M. Jorquera~Valero, P.~M. S{\'a}nchez~S{\'a}nchez, A.~Lekidis, J.~Fernandez~Hidalgo, M.~Gil~P{\'e}rez, M.~S. Siddiqui, A.~Huertas~Celdran, and G.~Mart{\'\i}nez~P{\'e}rez, ``{Design of a Security and Trust Framework for 5G Multi-domain Scenarios},'' \emph{Journal of Network and Systems Management}, vol.~30, no.~1, p.~7, 2022.

\bibitem{DIDSSI6G}
S.~Rodriguez~Garzon, H.~Yildiz, and A.~Küpper, ``{Decentralized Identifiers and Self-sovereign Identity in 6G},'' \emph{Trust, Security and Privacy of 6G of IEEE Networks}, 2021.

\bibitem{garzon2024beyond}
S.~R. Garzon, H.~Dinh-Tuan, M.~M. Martinez, A.~K{\"u}pper, H.~J. Einsiedler, and D.~Schneider, ``Beyond certificates: 6g-ready access control for the service-based architecture with decentralized identifiers and verifiable credentials,'' in \emph{2024 Joint European Conference on Networks and Communications \& 6G Summit (EuCNC/6G Summit)}.\hskip 1em plus 0.5em minus 0.4em\relax IEEE, 2024, pp. 830--835.

\bibitem{WorldWideWebConsortium.822021}
\BIBentryALTinterwordspacing
{World Wide Web Consortium (W3C)}, ``{Decentralized Identifiers v1.0},'' accessed: 2023-10-25. [Online]. Available: \url{https://www.w3.org/TR/did-core/}
\BIBentrySTDinterwordspacing

\bibitem{DIDCommAries}
\BIBentryALTinterwordspacing
D.~Hardman, ``{Aries RFC 0005: DID Communication},'' 2019, accessed: 2023-10-25. [Online]. Available: \url{https://github.com/hyperledger/aries-rfcs/blob/main/concepts/0005-didcomm/README.md}
\BIBentrySTDinterwordspacing

\bibitem{iyengar2021rfc}
J.~Iyengar and M.~Thomson, ``{RFC 9000: QUIC: A UDP-based multiplexed and secure transport},'' \emph{Internet Engineering Task Force}, 2021.

\bibitem{cho2020twinpeaks}
E.~Cho, J.~Kim, M.~Park, H.~Lee, C.~Hamm, S.~Park, S.~Sohn, M.~Kang, and T.~T. Kwon, ``{TwinPeaks: An approach for certificateless public key distribution for the internet and internet of things},'' \emph{Computer Networks}, vol. 175, p. 107268, 2020.

\bibitem{dahlmanns2022missed}
M.~Dahlmanns, J.~Lohm{\"o}ller, J.~Pennekamp, J.~Bodenhausen, K.~Wehrle, and M.~Henze, ``{Missed opportunities: Measuring the untapped TLS support in the industrial Internet of Things},'' in \emph{{Proceedings of the 2022 ACM on Asia Conference on Computer and Communications Security}}, 2022, pp. 252--266.

\bibitem{varo2022dynamic}
Q.~Varo, W.~Lardier, and J.~Yan, ``{Dynamic Reduced-Round TLS Extension for Secure and Energy-Saving Communication of IoT Devices},'' \emph{IEEE Internet of Things Journal}, vol.~9, no.~23, pp. 23\,366--23\,378, 2022.

\bibitem{selander2021ephemeral}
G.~Selander, J.~Mattsson, and F.~Palombini, ``{Ephemeral Diffie-Hellman over COSE (EDHOC)},'' \emph{{Internet Engineering Task Force, Internet-Draft draft-ietf-lake-edhoc-09}}, 2021.

\bibitem{jan2022selwak}
S.~A. Jan, N.~U. Amin, J.~Shuja, A.~Abbas, M.~Maray, and M.~Ali, ``{SELWAK: A Secure and Efficient Lightweight and Anonymous Authentication and Key Establishment Scheme for IoT Based Vehicular Ad hoc Networks},'' \emph{{Sensors}}, vol.~22, no.~11, p. 4019, 2022.

\bibitem{theodore2021novel}
S.~K.~A. Theodore, K.~R. Gandhi, and V.~Palanisamy, ``A novel lightweight authentication and privacy-preserving protocol for vehicular ad hoc networks,'' \emph{Complex \& Intelligent Systems}, pp. 1--11, 2021.

\bibitem{WorldWideWebConsortium.05.11.2021}
\BIBentryALTinterwordspacing
{World Wide Web Consortium (W3C)}, ``{Verifiable Credentials Data Model v2.0},'' accessed: 2023-10-25. [Online]. Available: \url{https://www.w3.org/TR/vc-data-model-2.0/}
\BIBentrySTDinterwordspacing

\bibitem{ETSI}
{European Telecommunications Standards Institute (ETSI)}, ``{Procedures for the 5G System (5GS)},'' \emph{3GPP TS 23.502 version 17.4.0 Release 17}, 2022.

\end{thebibliography}
\end{document}